# Transmission eigenchannels and the densities of states of random media

Matthieu Davy,[1,2*] Zhou Shi,[1*] Jing Wang,[1] Xiaojun Cheng[1] and Azriel Z. Genack[1]

[1] *Department of Physics, Queens College of the City University of New York, Flushing, NY 11367, USA.*
[2] *Institut d'Electronique et de Télécommunications de Rennes, University of Rennes 1, Rennes 35042, France.*
[*] *These authors contributed equally to this work*

We show in microwave measurements and computer simulations that the contribution of each eigenchannel of the transmission matrix to the density of states (DOS) is the derivative with angular frequency of a composite phase shift. The accuracy of the measurement of the DOS determined from transmission eigenchannels is confirmed by the agreement with the DOS found from the decomposition of the field into modes. The distribution of the DOS, which underlies the Thouless number, is substantially broadened in the Anderson localization transition. We find a crossover from constant to exponential scaling of fluctuations of the DOS normalized by its average value. These results illuminate the relationships between scattering, stored energy and dynamics in complex media.

The transmission matrix (TM) is the basis of a powerful approach to quantum and classical wave propagation that is able to explain the statistics of conductance [1-7] and transmission [8] and prescribe the degree to which the transmitted wavefront can be manipulated [9-15]. It was introduced by Dorokhov to explain the scaling of electronic conductance in wires[1]. For a wire connected to perfectly conducting leads that support $N$ propagating channels, and similarly for a waveguide between segments of empty waveguides, the $N \times N$ elements of thee TM, $t$, $t_{ba}$, are the field transmission coefficients through the sample between the incident channels $a$ and outgoing channels $b$. The conductance in units of the quantum of conductance $G/(e^2/h)$ is equivalent to the classical transmittance $T$ which can be expressed in terms of the transmission eigenvalues $\tau_n$ of the matrix product $tt^\dagger$, $T = \sum_{n=1}^{N} \tau_n$ [1, 6]. The probability distribution of these transmission eigenvalues determines the statistics of transmission [1-5, 8, 16].

For diffusive samples, the average of $T$ over a random ensemble is given by $\langle T \rangle \equiv g = \xi/L$, where $\xi = N\ell$ is the localization length and $\ell$ is the transport mean free path.[17] For $g>1$, transport is diffusive and the flux transmitted through disordered samples in eigenchannels of transmission varies over a wide range with a small number of highly transmissive channels among a multitude of dark eigenchannels[1-5]. The transmittance is dominated by the approximately $g$ "open" channels with transmission eigenvalues $\tau_n>1/e$.

Recently, considerable attention has focused on the power of the TM to mold the flow of waves through random samples [14] and to modify the energy density inside the medium [18, 19]. The possibility of sharp focusing and enhanced transmission has been demonstrated for sound [20], elastic waves [21], light [9, 13, 22] and microwave radiation [12]. These phenomena have been described in terms of the transmission eigenvalues [9-13, 21], but static transmission parameters cannot explain the dynamics of transmission or provide the DOS whose statistics control emission, absorption and wave localization and give the proclivity of a medium to emit radiation and store energy [23-33]. The crossover to wave localization reflects the changing character of the underlying modes, from extended to spatially peaked. This is characterized by the Thouless number δ, which is essentially the ratio of the typical linewidth to the spacing of classical modes or quantum energy levels [17, 34]. But the full statistics of modal overlap, reflected in δ have not been measured.

In this Letter, we show that the DOS can be determined from measurements of spectra of the TM. The contribution of each eigenchannel to the DOS is the derivative with angular frequency of a composite phase of the eigenchannel. Summing the contributions from all eigenchannels provides the first direct measurement of the DOS of a multiply scattering medium as a whole. The DOS determined from the eigenchannels is found to be in excellent agreement with the DOS found from a decomposition of the transmitted field into modes. The probability distribution of the DOS broadens substantially in the crossover to Anderson localization reflecting the increasing spectral isolation of long-lived localized modes. The eigenchannel phase derivative, which is equal to the delay time in transmission, increases with $\tau_n$. When normalized by the average delay time, the eigenchannel delay time versus $\tau_n$ for diffusive samples of different length is found to fall on universal curve.

The DOS of a bounded open medium for classical waves is the density of quasi-normal modes or resonances of a region per unit angular frequency,[34]

$$\rho(\omega) = \frac{1}{\pi} \sum_n \frac{\Gamma_n/2}{(\Gamma_n/2)^2 + (\omega - \omega_m)^2}. \quad (1)$$

Here $\omega_n$ is the central frequency and $\Gamma_n$ is the linewidth of the $n^{th}$ mode. The integral over frequency of each mode in Eq. (1) is unity. Krein, Birman and Schwinger [24-26] have shown that the DOS may be expressed in terms of the scattering matrix $S$, $\rho(\omega) = -i\frac{1}{2\pi} Tr S^\dagger dS/d\omega$, where $-iS^\dagger dS/d\omega$ is the Wigner-Smith delay-time matrix whose trace is equal to the sum of scattering times in all 2N channels linked to the scattering region [27, 28]. This is proportional to the integral of the energy stored within the medium for unit incident flux in each channel, $\rho(\omega) \propto \sum_\alpha^{2N} \int_V I_\alpha(r,\omega) dV$ [28, 30]. The difficulty of carrying out measurements over all possible scattering channels has so far precluded a determination of the DOS based on the scattering matrix. However, the calculations of Brandbydge and Tsukada [29] of the local DOS of electrons based on the scattering matrix show that the DOS can be determined from measurements of the TM. The DOS can be obtained from the summation of the derivative of composite phase of the transmission eigenchannels with angular frequency. The phase derivative of the $n^{th}$ transmission eigenchannel, $\frac{d\theta_n}{d\omega} = \frac{1}{i}(\mathbf{u}_n^* \frac{d\mathbf{u}_n}{d\omega} - \mathbf{v}_n^* \frac{d\mathbf{v}_n}{d\omega})$, where $\mathbf{v_n}$ and $\mathbf{u_n}$ are $n^{th}$ columns of the unitary matrices $V$ and $U$, may be obtained from the singular value decomposition of the TM, $t = U\Lambda V^\dagger$ [35]. $\Lambda$ is a diagonal matrix with elements $\sqrt{\tau_n}$. The DOS is then

$$\rho(\omega) = \frac{1}{\pi} \sum_{n=1}^{N} \frac{d\theta_n}{d\omega}. \qquad (2)$$

Each term in the sum is the contribution of a single eigenchannel to the DOS, the eigenchannel density of states (EDOS).

This relation is an extension to multichannel systems of the equality between the DOS and the transmission delay time in 1D systems [31]. The eigenchannel phase derivative $d\theta_n/d\omega$ corresponds to the delay time $\Delta t_n(\omega)$ of energy for a transmitted pulse composed of a superposition of waves in the $n^{th}$ eigenchannel centered at $\omega$ in the limit of vanishing bandwidth [35].

Measurements of the TM for which the impact of absorption is removed are carried out in a copper waveguide containing randomly positioned alumina spheres [10, 16, 35, 36]. The empty waveguide supports $N \sim 66$ modes for diffusive waves and $N \sim 30$ for localized waves. Measurements are made for linearly polarized horizontal and vertical components of the field over the front and back surfaces of the waveguide by translating and rotating wire antennas on a square grid. The TM is computed using $N/2$ points for each polarization for diffusive wave. For localized waves, the measurements reported here are made for only a single polarization. The measurement of the TM on a grid for a single polarization is incomplete [11, 13], but we find that the statistics of the TM are well represented by the measured TM provided that the measured size of the TM, $N'$, is much greater than the value of dimensionless conductance $g$ in the sample [11, 16]. New configurations are obtained by briefly rotating and vibrating the sample tube. For diffusive waves, the TM was measured for three sample lengths, $L=23$, 40 and 61 cm, while for localized waves, measurements are reported here for samples of length $L=40$ cm.

Microwave spectra of $\tau_n$ and $d\theta_n/d\omega$ for a single random configuration drawn from an ensemble of diffusive samples with $g=6.9$ and localized samples with $g=0.37$ are shown in Fig. 1. In Fig. 2 we compare the DOS determined from the sum in Eq. (1) of contributions from all modes to the DOS given by the sum over eigenchannels in Eq. (2). The comparison is made for waves in the crossover to Anderson localization for which the degree of modal overlap is appreciable but still small enough that the full set of mode central frequencies $\omega_n$ and linewidths $\Gamma_n$ and so the contribution to the DOS for each mode can be accurately determined from a decomposition of field spectra [37]. The DOS found from the modal decomposition involves the analysis of the entire field spectrum and modes can be found from measurements of the TM as well as from measurements of field spectra within the interior of the sample [35], from which it is impossible to find the transmission eigenchannels. In contrast, the analysis of the transmission eigenchannels requires only the TM at two slightly shifted frequencies so that the derivative of the phase can be found. Thus the DOS determined from an analysis of modes and channels is independent. A plot of the spectrum of the individual modes corresponding to the terms in Eq. (1) is shown in Fig. 2a. Good agreement is found in Fig. 2b between the sums of the contributions to the DOS of all eigenchannels and of all modes determined from the TM and from spectra of the field inside the sample. The analysis of the TM can thus be used to find the DOS in samples with strong modal overlap for modal analysis is not possible.

The degree of overlap of the modes of the medium is a fundamental indicator of the nature of wave propagation. This is encapsulated in the Thouless number, which is essentially the average of the ratio of the spectral width and spacing of modes [17, 37]. A more comprehensive representation of the nature of modal overlap in random systems would be the probability distribution of the relative DOS, $\rho(\omega)/<\rho(\omega)>$, which can be constructed from spectra of $\sum_{n=1}^{N} d\theta_n / d\omega$. The probability distributions $P(\sum_{n=1}^{N} \frac{d\theta_n}{d\omega} / \langle \sum_{n=1}^{N} \frac{d\theta_n}{d\omega} \rangle)$ measured for ensembles of samples with $L_{eff}/\xi$ ranging from 0.14 to 2 are shown in Fig. 3a. The effective sample length [38] is $L_{eff} = L + 2z_b$ with intensity extrapolation lengths beyond the sample reflecting internal reflection of $z_b$=6 cm for localized waves and 13 cm for diffusive waves [10]. The distributions are seen to broaden with increasing sample length $L$, particularly beyond the crossover to localization at $L/\xi =1$ when distinct peaks begin to emerge in the spectrum of the DOS. For $L_{eff}/\xi = 2.08$, the probability distribution of the DOS, $P(\rho)$, is seen in Fig. 3b to exhibit an algebraic tail as $1/\rho^{4.8}$ in agreement with simulations for this sample [35]. For $L >> \xi$, the tail of the time delay distribution for transmitted waves in a random 1D sample, which is the same as the DOS, is $P(\rho) \propto 1/\rho^2$ [39, 40]. The probability distribution of transmittance, $P(T)$, for quasi-1D samples is found to approach the log-normal distribution predicted for $L >> \xi$ in 1D samples [41] only when the participation number of transmission eigenchannels[12], $M \equiv (\sum_{n=1}^{N} \tau_n)^2 / \sum_{n=1}^{N} \tau_n^2$ is very close to unity. $P(T)$ for the sample with $L_{eff}/\xi$=2.08 is a one-sided log-normal distribution [16] and we would not expect $P(\rho)$ to reach its asymptotic form for $L/\xi >> 1$.

We carry out two-dimensional numerical simulations using the recursive Green's function method [42] to explore the fluctuations of the DOS as well as to determine the impact of an incomplete measurement of the TM on estimates of the DOS. Spectra of $d\theta_n / d\omega$ for a diffusive sample with $L/\xi = 0.69$ and $N$=33 in which fluctuations in spectra of the EDOS are still appreciable are shown in Fig. 4a. $d\theta_n / d\omega$ is seen to coincide with normalized spectra of the integral of the energy density over the sample volume for a transmission eigenchannel $\int_V I_n(r,\omega) dV$ [35], confirming that $d\theta_n / d\omega$ is the contribution of the $n^{th}$ eigenchannel to the DOS, the EDOS. However, it is difficult to excite and detect all channels and the measured TM for diffusive waves is typically incomplete. $(d\theta_n / d\omega)/\pi$ is then no longer equal to the EDOS and its sum does not give the DOS. Similarly, it is not possible to construct a fully transmitted incident wave when the TM is not complete [11]. We estimate that the best agreement between measurements of $\langle d\theta_n / d\omega \rangle$ and simulations of $\langle d\theta_n / d\omega \rangle$ shown in Fig. 5b occurs when we construct the simulated TM using a fraction $N'/N = 0.7$ of the channels of the system. The best agreement between measurements and calculations of the probability distribution of transmission eigenvalues is also obtained for $N'/N = 0.7$ [35].

In Fig. 4a slight differences between spectra of the sum of $d\theta_n / d\omega$ for a complete TM and for a sample with $N'/N = 0.7$ are observed. The probability distributions $P(\sum_{n=1}^{N} d\theta_n / d\omega)$ shown in

Fig. 4b are quite similar when $N' \gg M$. Significant deviations arise, however, for diffusive samples in which $M$ is comparable to or larger than $N'$. The number of measured channels is then insufficient to accurately reflect the nature of transport.

In Fig. 4b and 4c, we show the results of simulation of the scaling of the variance of $\rho(\omega)/<\rho(\omega)>$ for samples with $N$=16 and $N$=33 together with the result obtained from the distribution shown in Fig. 3 for $L_{eff}/\xi = 2.07$. In this case $N' \gg M$. Measurements are seen to be in good agreement with simulations. The variance for diffusive waves, for $L/\xi < 1$, is flat with a value of $\sim 0.003$ for $N$=33 and $\sim 0.007$ for $N$=16. Rigidity in the spectrum of the central frequencies of electromagnetic modes when many modes fall within the mode linewidth [43] is likely the origin of the constant value of the variance of $\rho(\omega)/<\rho(\omega)>$ for each value of $N$. For deeply localized waves, mode spacing typically exceeds the linewidth so that fluctuations of the DOS increases rapidly with $L/\xi$ as modal linewidths fall. The variance of $\rho(\omega)/<\rho(\omega)>$ is seen in Fig. 4c to increases exponentially as $e^{1.6L/\xi}$. Thus fluctuations of the DOS provide an experimental measure of $L/\xi$ and of the degree of modal overlap for both diffusive and localized waves [43].

The variation of the transmission delay time with the transmission eigenvalue in diffusive samples is shown in Fig. 5a. $\langle d\theta_n/d\omega \rangle$ increases with $\tau_n$ and sample length so that delay times are lengthened in coherent eigenchannels with high transmission. When normalized by the ensemble average of the photon delay time, which is the average of the single channel delay time between channels a and b weighted by the transmitted intensity $|t_{ba}|^2$, $\langle d\varphi/d\omega \rangle = <\sum_{ab}|t_{ab}|^2 d\varphi_{ab}/d\omega>/<\sum_{ab}|t_{ab}|^2>$, the measurements collapse to a single curve (Fig. 5b). This is confirmed in simulations as shown in Fig. 5c. The constant ratio of the delay time $\langle d\theta_n/d\omega \rangle$ averaged over eigenchannels of fixed $\tau_n$, to the average delay time together with the scaling of $\langle d\varphi/d\omega \rangle$ as $L^2$ for diffusive waves [44] indicates that the EDOS for a given value of $\tau_n$ scales as $L^2$, as seen in Fig. 5d for $\tau_n$=0.1 and 1. Though the EDOS scales as $L^2$, the DOS is seen in Fig 5d to scale as $L$, as expected, since the number of open channels with $\tau_n > 1/e$ is proportional to $g = \xi/L$ [1, 2], falls inversely with $L$.

We have shown that it is possible to measure the dynamics and stored energy in addition to the transmission of each transmission eigenchannel. This makes it possible to measure the DOS as well as the transmittance for both diffusive and localized waves. Fluctuations in the DOS can provide a rich picture of the changing nature of transport in the Anderson localization transition. The cumulant correlation function with frequency shift of the DOS normalized to its average may yield the statistics of the spacing of energy levels and the probability of return of scattered particle of the wave to a coherence volume within the sample vs. time delay. The wave becomes localized when the probability of return integrated over time equals unity [43, 45]. Selective excitation of highly-transmissive, long-lived eigenchannels can enhance energy collection and lower the threshold of random lasers.

We thank Evgeni Gurevich for stimulating discussion on the scattering matrix and Arthur Goetschy and A. Douglas Stone for providing the simulation code to calculate the transmission matrix through a two dimensional disordered waveguide. The research was supported by the National Science Foundation (DMR-1207446) and by the Direction Générale de l'Armement (DGA).

**FIGURES**

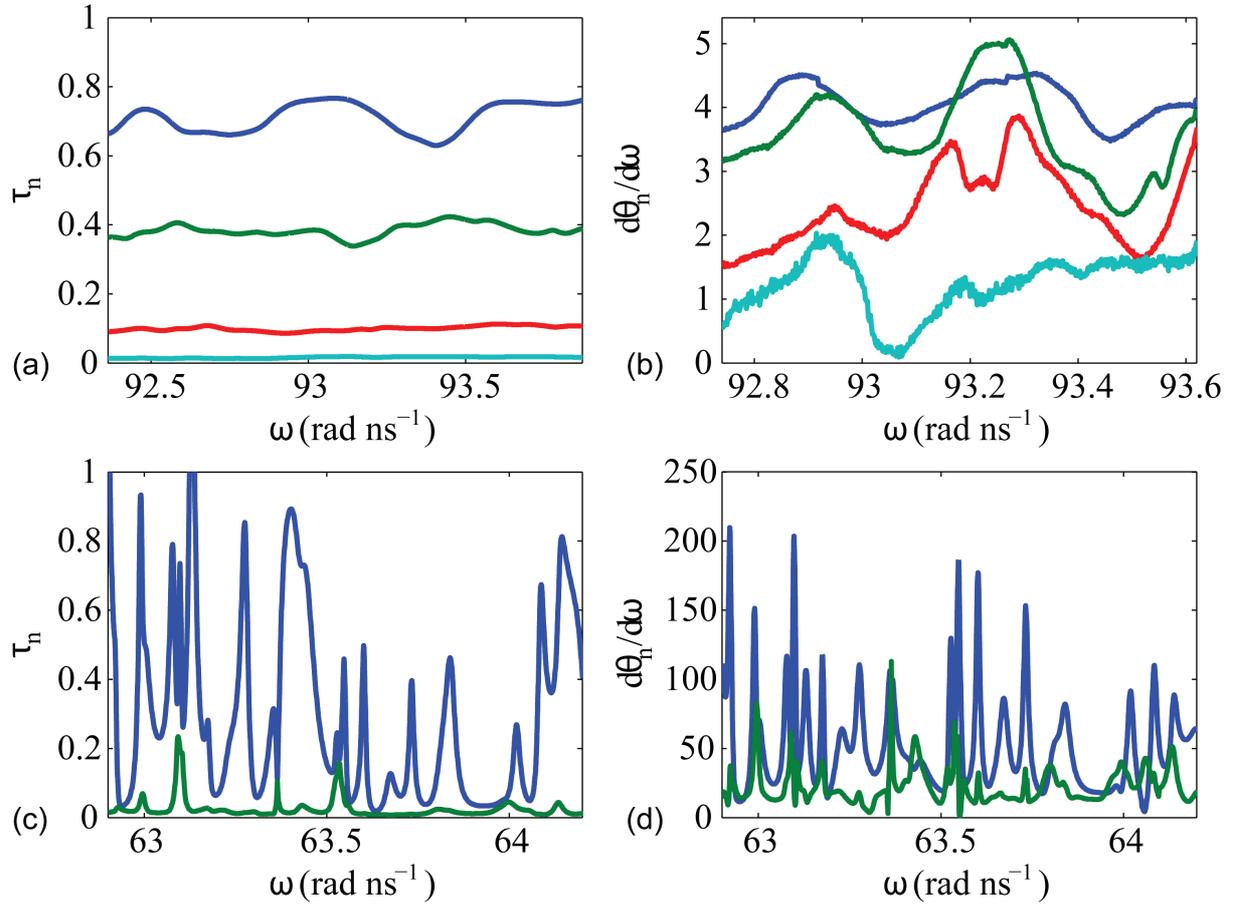

FIG. 1 (color online). Spectra of $\tau_n$ (a,c) and spectra of $(d\theta_n/d\omega)$ (b,d) for eigenchannels $n=1,5,15,25$ for diffusive waves of sample length $L=23$ cm with $g=6.9$ (a,b), and n=1,2 for localized waves of length $L=40$ cm and $g=0.37$ (c,d).

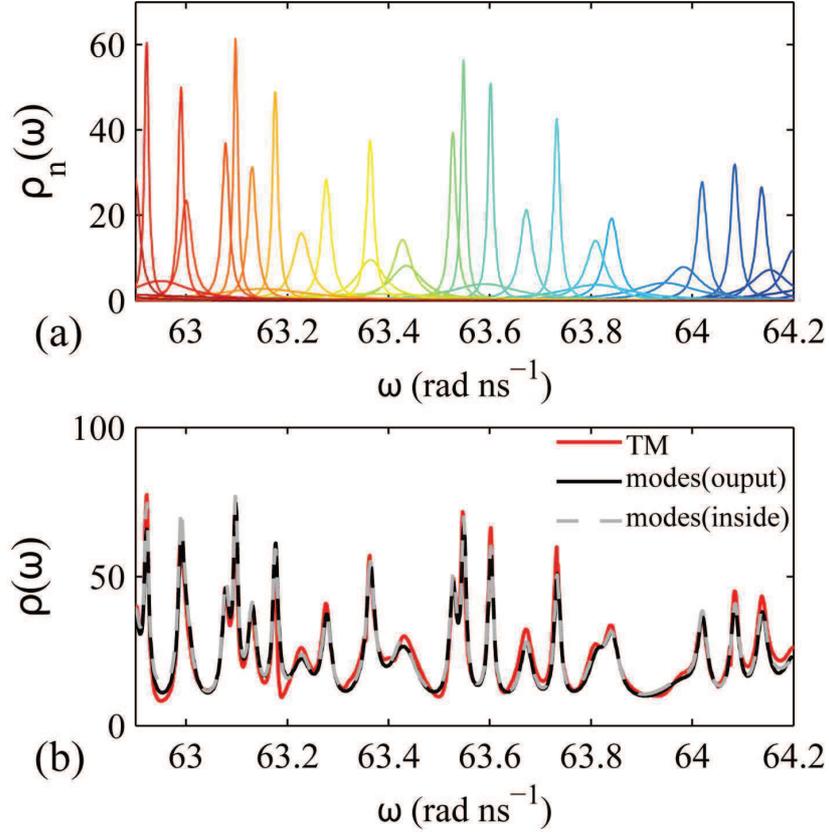

FIG. 2 (color online). (a) Contributions of the individual modes in Eq. (1) to the DOS. (b) Comparison of the DOS determined from the TM (red curve) by summing spectra of $(d\theta_n/d\omega)$ and modes found from spectra of the field at the output (black curve) and from spectra of the field inside the sample (grey dotted curve).

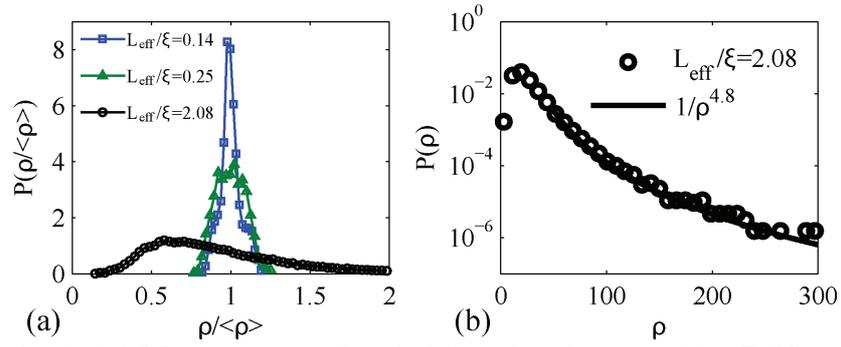

FIG. 3 (color online). (a) Measurement of probability distribution of the DOS normalized by its average value determined from the sum of $d\theta_n/d\omega$ for $L_{eff}/\xi=0.14$ (blue squares), $L_{eff}/\xi=0.25$ (green triangles) and $L_{eff}/\xi=2.08$ (black circles). (b) Probability distribution of the DOS for $L_{eff}/\xi=2.08$ in a semilog scale. The black line is a fit of the tail as $1/\rho^{4.8}$.

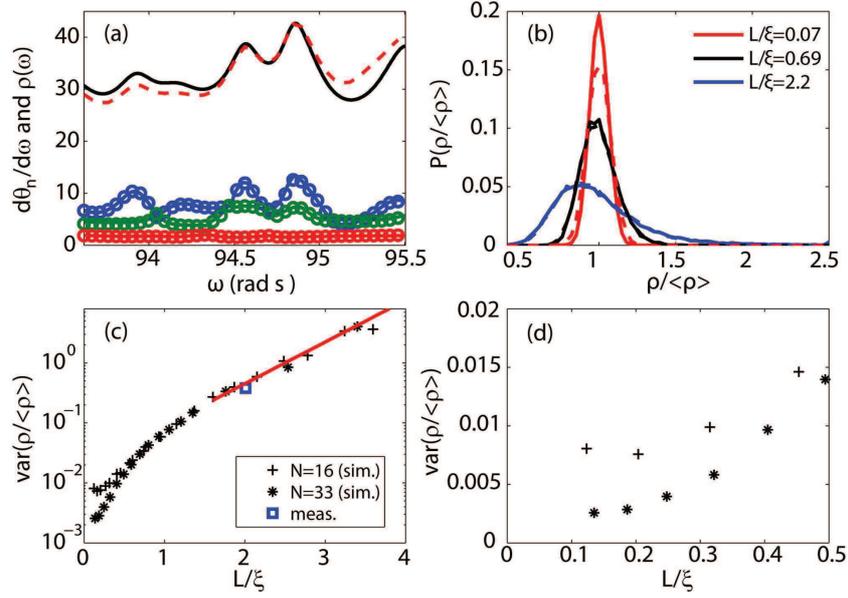

FIG. 4 (color online). (a) Spectra of $d\theta_n/d\omega$ (full curves) and the normalized integral of the energy density $I_n(z)$ inside the sample (circles), for eigenchannels with $\tau = 0.8$ (blue line), $\tau = 0.5$ (green line), $\tau = 0.01$ (red line) for $N=33$ and $L/\xi=0.69$. The black curve is $\rho(\omega)$. The red dashed curve is $\sum_{n=1}^{N'} d\theta_n/d\omega$ for an incomplete measurement of the TM with $N'/N=0.7$. (b) Probability distribution $P(\rho/<\rho>)$ for $L/\xi = 0.07$ (red line), $L/\xi = 0.69$ (black line) and $L/\xi = 2.2$ (blue line). The corresponding dashed curves are $P(\sum_{n=1}^{N'} d\theta_n/d\omega/<\sum_{n=1}^{N'} d\theta_n/d\omega>)$ with $N'/N = 0.7$. (c) Variation of the variance of the normalized DOS, $\text{var}(\rho/<\rho>)$, as a function of $L/\xi$ in simulation with $N=16$ (black crosses) and $N=33$ (black stars) and in measurements (squares). The red line is an exponential fit of the data for localized waves, $L/\xi > 1$. The results are obtained from 5000 simulations of samples with the same length but different degree of disorder. (d) Linear plot of $\text{var}(\rho/<\rho>)$ for diffusive waves.

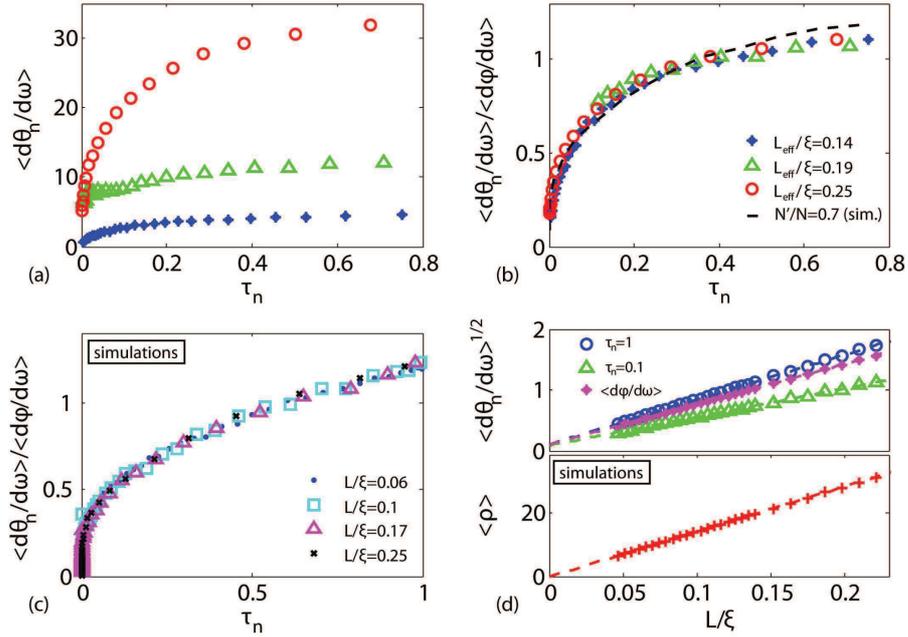

FIG. 5 (color online). (a) Averages of measured $\langle d\theta_n/d\omega\rangle$ and (b) $\langle d\theta_n/d\omega\rangle/\langle d\varphi/d\omega\rangle$, vs. $\tau_n$ for $L$=61cm (red circles), $L$=40 cm (green triangles) and $L$=23 cm (blue filled circles). The black curve is obtained from simulations in which $N'/N = 0.7$. (c) simulations for the complete TM of $\langle d\theta_n/d\omega\rangle/\langle d\varphi/d\omega\rangle$ with $L/\xi = 0.06$ (blue dots), $L/\xi = 0.1$ (cyan squares), $L/\xi = 0.17$ (magenta triangles) and $L/\xi = 0.25$ (black crosses). (d) Scaling of $\sqrt{\langle d\theta_n/d\omega\rangle}$ for eigenchannels with $\tau = 1$ (blue dots) and $\tau = 0.1$ (blue triangles), of $\sqrt{\langle d\varphi/d\omega\rangle}$ (blue stars), and of $<\rho>$ (red crosses).

# Supplementary Information for 'Transmission eigenchannels and the densities of states of random media'

Matthieu Davy,[1,2] Zhou Shi,[1] Jing Wang,[1] and Azriel Z. Genack[1]

[1]Department of Physics, Queens College of the City University of New York, Flushing, NY 11367, USA.
[2]Institut d'Electronique et de Télécommunications de Rennes, University of Rennes 1, Rennes 35042, France.

## 1. Derivation of the density of states in terms of eigenchannel dwell times

Krein, Birman and Schwinger [1-5] have shown that the density of states (DOS) is given as,

$$\rho(\omega) = -i\frac{1}{2\pi}TrS^\dagger dS/d\omega \qquad (1)$$

Here $S$ is the scattering matrix consisting of four $N \times N$ matrices, in transmission and reflection for waves incident on the two open boundaries of the sample, $S = \begin{pmatrix} r & t' \\ t & r' \end{pmatrix}$. The unprimed matrices are for the wave coming in on the left and the primed matrices indicate the wave is incident on the right side of the sample. The DOS can thus be expressed in terms of the components of the scattering matrix,

$$-\frac{i}{2\pi}Tr(S^\dagger(\omega)\frac{dS(\omega)}{d\omega}) = -\frac{i}{2\pi}Tr(r^\dagger\frac{dr}{d\omega} + t^\dagger\frac{dt}{d\omega} + t'^\dagger\frac{d(t')}{d\omega} + r'^\dagger\frac{d(r')}{d\omega}) \qquad (2)$$

Following the approach of Brandbydge and Tsukada [6] who calculated the local DOS of electrons in terms of the derivative of the phase of the transmission eigenchannel with respect to the change of local potential, we show that the DOS can be obtained from the measurement of the TM, for a non-absorbing system with time-reversal symmetry.

Via the singular value decomposition of the TM, $t = U\Lambda V^\dagger$, $Tr(t^\dagger \frac{dt}{d\omega})$ is given by,

$$-iTr(t^\dagger \frac{dt}{d\omega}) = \text{Im}[Tr(\Lambda^2 U^\dagger \frac{dU}{d\omega} + \Lambda \frac{d\Lambda}{d\omega} + \Lambda^2 \frac{dV^\dagger}{d\omega}V)] \qquad (3)$$

Because $\Lambda$ is a real matrix,

$$-iTr(t^\dagger \frac{dt}{d\omega}) = \text{Im}[Tr(\Lambda^2 U^\dagger \frac{dU}{d\omega} + \Lambda^2 \frac{dV^\dagger}{d\omega}V)] \qquad (4)$$

Since, $V$ is a unitary matrix, $VV^\dagger = I$, this yields $\frac{dV^\dagger}{d\omega}V = -V^\dagger \frac{dV}{d\omega}$. Eq. (3) can then be written as,

$$-iTr(t^\dagger \frac{dt}{d\omega}) = -iTr(\Lambda^2 U^\dagger \frac{dU}{d\omega} - \Lambda^2 V^\dagger \frac{dV}{d\omega}) \qquad (5)$$

Finally, this leads to,

$$-iTr(t^\dagger \frac{dt}{d\omega}) = -i\sum_n \tau_n(\mathbf{u}_n^* \frac{d\mathbf{u}_n}{d\omega} - \mathbf{v}_n^* \frac{d\mathbf{v}_n}{d\omega}) = \sum_{n=1}^{N} \tau_n \frac{d\theta_n}{d\omega} \qquad (6)$$

Where, $\frac{d\theta_n}{d\omega} = \frac{1}{i}(\mathbf{u}_n^* \frac{d\mathbf{u}_n}{d\omega} - \mathbf{v}_n^* \frac{d\mathbf{v}_n}{d\omega})$.

Because of the time-reversal symmetry $t = (t')^T$, the contributions from the two transmission matrices are the equal,

$$-iTr(t'^\dagger \frac{dt'}{d\omega}) = \sum_{n=1}^{N} \tau_n \frac{d\theta_n}{d\omega} \qquad (7)$$

Due to the conservation of flux, the eigenvalues of the $rr^\dagger$ and $r'r'^\dagger$ are, $1-\tau_n$ and the matrices $rr^\dagger$ and $r'r'^\dagger$ can be written with the unitary matrices $U$ and $V$, $rr^\dagger = V^*(I-\tau)V^T$ and $r'r'^\dagger = U(I-\tau)U^\dagger$. Following the same procedure as above, we obtain,

$$-iTr(r^\dagger \frac{dr}{d\omega} + r'^\dagger \frac{dr'}{d\omega}) = 2(1-\tau_n)\frac{d\theta_n}{d\omega} \qquad (8)$$

Summing Eq. (6-8) gives the DOS,

$$-\frac{i}{2\pi}Tr(S^\dagger(\omega)\frac{dS(\omega)}{d\omega}) = \frac{1}{\pi}\sum_{n=1}^{N}\frac{d\theta_n}{d\omega} \qquad (9)$$

Equation (9) shows that $(d\theta_n/d\omega)/\pi$ is the contribution of the $n^{th}$ eigenchannel to the DOS.

In Fig. S1 we explore the relationship between the integral of the energy density inside the sample for the $n^{th}$ eigenchannel, denoted by $\int_V I_n(r,\omega)dV$ and $d\theta_n/d\omega$ with computer simulations. We consider a scalar wave within a two-dimensional disordered waveguide with semi-infinite ideal leads with index of refraction of unity and perfectly reflecting transverse sides. The disordered region within a multimode waveguide is modeled by a random position-dependent dielectric constant $\varepsilon(x,y)$. The wave equation $\nabla^2 E(x,y) + k_0^2 \varepsilon(x,y)E(x,y) = 0$ is discretized on a square gird. Here, $k_0$ is the wave number in the sample leads. The Green's function at points on the grid is calculated via the recursive Green's function method [7].

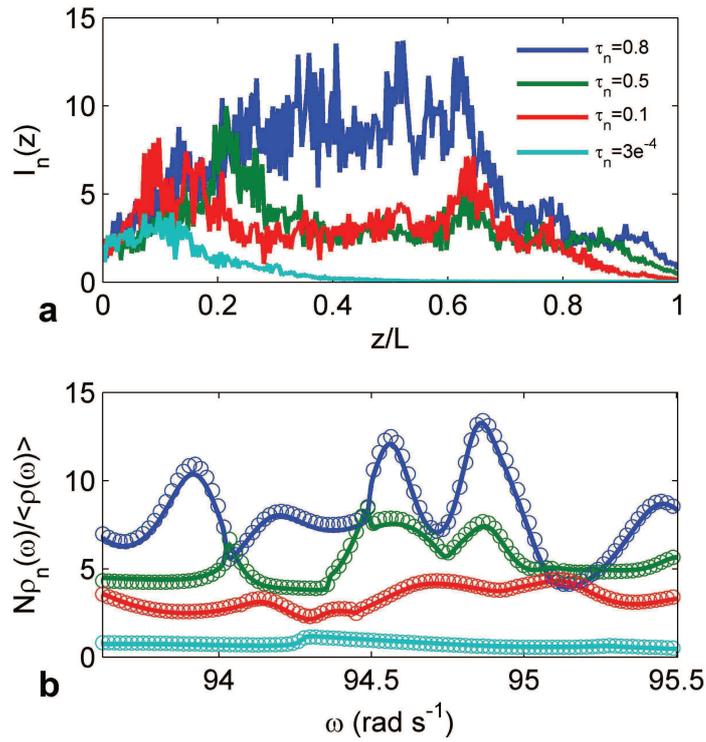

**Figure S1 | Relationship between dwell times and DOS.** (a) Intensity inside the sample for eigenchannels with $\tau_n$=0.8 (blue line), $\tau_n$=0.5 (green line), $\tau_n$=0.1 (red line) and $\tau_n$ =0.001 (cyan line). for $N$=33 and $L/\xi$=0.69. (b) Corresponding spectra of the eigenchannels delay times (circles) and the EDOS (curves).

We choose a diffusive sample with $L/\xi$=0.69 so that fluctuations in spectra of $d\theta_n/d\omega$ are clearly observed. The variation of intensity integrated over the sample cross section versus depth $z$ into the sample is shown in Fig. S1a for a number of transmission eigenchannels. Plots of $\int_V I_n(r,\omega)dV$ and $d\theta_n/d\omega$ normalized to the average over channels are presented in Fig. S1b and are seen to coincide. These results confirm that $d\theta_n/d\omega$ is also proportional to the energy stored in the $n^{th}$ eigenchannel. Just as the sum of $\tau_n$ gives $T$, the sum of $(d\theta_n/d\omega)/\pi$ gives the DOS showing that propagation in multichannel samples can be regarded as the sum of characteristics of independent 1D systems.

## 2. Measurement of the TM and determination of the modes of a random system using the spectra of the field inside the sample

Measurements of the TM are carried out in a copper waveguide of diameter 7.3 cm containing randomly positioned alumina spheres [8, 9]. The impact of absorption is removed by Fourier transforming field spectra to form a pulse into the time domain, multiplying the time dependent field by a factor $\exp(t/2\tau_a)$, where $1/\tau_a$ is the absorption rate, and finally Fourier transforming back into the frequency domain [10, 11]. Measurements are analyzed in two frequency ranges 14.7-14.94 GHz and 10-10.24 GHz in which the wave is diffusive and localized, respectively. Measurements are made for linearly polarized horizontal and vertical components of the field over the front and back surfaces of the waveguide by translating and rotating wire antennas on a square grid with spacing of 9mm. For diffusive waves, the TM was measured for three sample lengths, $L$=23, 40 and 61 cm for 23, 4 and 6 sample configurations, respectively, while for localized waves, measurements are reported here for samples of length $L$= 40 cm for 60 sample realizations.

The determinations of the DOS from the analyses of modes and transmission eigenchannels in Fig. 2 are in principle distinct and independent. The DOS found from the modal decomposition at each frequency involves the analysis of the entire spectrum of the Green's function, while the analysis of the transmission eigenchannels requires only the TM at two slightly shifted frequencies so that the derivative of the phase can be found. Field spectra from at any set of points within the medium could therefore be utilized as well to find the modes. This is demonstrated in Fig. 2 of the main text in comparisons of the DOS obtained from Eq. (1) with modal properties $\omega_n$ and $\Gamma_n$ found from a fit of the expression $E(r,r',\omega) = \sum_n a_n(r,r') \frac{\Gamma_n/2}{\Gamma_n/2 + i(\omega - \omega_n)}$ to measurements of the field at points *inside* the sample and a determination based on transmission eigenchannels utilizing Eq. (2) in the main text.

To determine the modes from the field inside the medium, the same $L$=40 cm copper tube and scattering medium used in all the measurements reported here except for five 3-mm-diameter holes in a line along the length of the tube separated by 1 cm from 35-39 cm. The TM was measured with 45 input points and 45 output points. The modes were determined from the field transmission coefficients between a source antenna at various points on the input surface and a detection antenna moved to the center of each of the five holes. The DOS spectra of modes determined from measurement of the microwave field along the length of the sample and of channels determined from measurements through the sample are seen in Fig. 2b of the main text to be in reasonable agreement.

## 3. Relation between eigenchannel phase derivative and eigenchannel delay time

The DOS given by Eq. (2) can alternatively be expressed in terms of single channel delay times $d\varphi_{ab}/d\omega$, weighted by the scattered intensity $I_{ab}$ over the $2N$ incoming and outgoing channels on both sides of the sample [12, 13],

$$\rho(\omega) = \frac{1}{\pi} \sum_{a,b}^{2N} I_{ab} \frac{d\varphi_{ab}}{d\omega}. \tag{10}$$

The single channel delay times is the time spent in the sample by a wave injected into channel *a* and emerging in channel *b* in the limit of vanishing pulse bandwidth [14-16],

$$\Delta t_{ab}(\omega_0) = \lim_{\Delta\omega \to 0} \int I_{ab}(t;\omega_0,\Delta\omega)t\,dt \,/\, \int I_{ab}(t;\omega_0,\Delta\omega)dt = \frac{d\varphi_{ab}}{d\omega}\Big|_{\omega_0} \quad (11)$$

Here $I_{ab}(t;\omega_0,\Delta\omega) = |t_{ab}(t;\omega_0,\Delta\omega)|^2$ is the transmitted pulse with carrier frequency $\omega_0$ and bandwidth $\Delta\omega$. The time varying field $t_{ab}(t;\omega_0,\Delta\omega)$ is the inverse Fourier transform of the field spectrum $t_{ab}(\omega) = \sqrt{I_{ab}}e^{i\varphi_{ab}}$ with $I_{ab} = |t_{ab}|^2$.

In this section, we demonstrate that the delay time of an eigenchannels $\Delta t_n(\omega_0,\Delta\omega)$ for a pulse with vanishing bandwidth is the composite eigenchannel phase derivative $d\theta_n/d\omega$. We define $v_{na}(t)$ as the component in channel $a$ of the time-dependent wave composed from a narrow bandwidth of singular vector components $v_{na}(\omega)$ of unit incident flux in $v_n(\omega)$. $v_{na}(t)$ is the inverse Fourier transforms of $v_{na}(\omega)G(\omega)$, where $G(\omega)$ is an envelope with a narrow bandwidth. An example of $v_{na}(t)$ and the corresponding waveforms $u_{nb}(t) = FT^{-1}\left[\lambda_n(\omega)u_{nb}(\omega)G(\omega)\right]$ are presented in the inset of Fig. S2 for a Gaussian envelope.

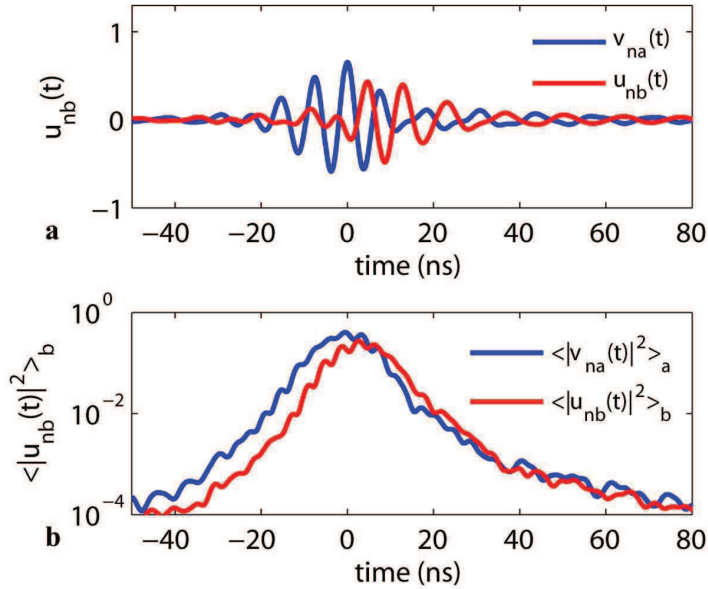

**Figure S2 | Pulse transmission for transmission eigenchannel determined from measured spectra of the transmission eigenchannel in a single random realization of the sample.** (a) plot of the time variation of the incident and outgoing field for the $n=1$ transmission eigenchannel for which the spectrum of $d\theta_n/d\omega$ is shown in Fig. 1a of the main text. The blue curve represents the incident wave and the red curve the outgoing wave. The sample configuration is drawn from a sample with g=6.9. The field variation is obtained from the inverse Fourier transforms $v_{na}(\omega)$ (blue curve) and $\lambda_n(\omega)u_{na}(\omega)$ (red curve). (b) Semilog-plot of the intensity variation for the $n=1$ transmission eigenchannel averaged over all channels.

The single-channel delay time in Eq. (11) gives the decay in transmission for an incident pulse centred at $t=0$. For the eigenchannel delay time $\Delta t_n(\omega_0,\Delta\omega)$, the incident pulse reflects the spectrum of the $n^{th}$ transmission eigenchannel and the time delay reflects the difference between the outgoing and incident pulse for this transmission eigenchannel. The eigenchannel delay time is then the average of the time delays on the outgoing and incident sides of the sample averaged over all output channels $a$ and output channels $b$ weighted by the corresponding fluxes

$$\Delta t_n(\omega_0, \Delta\omega) = <\frac{\int |u_{nb}(t;\omega_0,\Delta\omega)|^2 t\, dt}{\int |u_{nb}(t;\omega_0,\Delta\omega)|^2 dt}>_b - <\frac{\int |v_{na}(t;\omega_0,\Delta\omega)|^2 t\, dt}{\int |v_{na}(t;\omega_0,\Delta\omega)|^2 dt}>_a \quad (12)$$

Since the incident pulse $v_{na}(t)$ is not centered at t=0 the time corresponding to its barycenter is subtracted from the delay time associated to $v_{nb}(t)$. In the limit of vanishing pulse bandwidth, $\lambda_n(\omega)$ is constant over the bandwidth so that $FT^{-1}[\lambda_n(\omega)u_{nb}(\omega)G(\omega)] \simeq \lambda_n FT^{-1}[u_{nb}(\omega)G(\omega)]$ and the frequency variation of the singular values need not be considered.

The derivative of the phase of $u_{nb}(\omega)$, $d\varphi_{nb}^{(u)}/d\omega$, is given by $\frac{d\varphi_{nb}^{(v)}}{d\omega} = \frac{1}{u_{nb}} \frac{du_{nb}}{d\omega}$. The singular vectors are random variables with statistically independent real and imaginary parts and are normalized so that $\sum_a |v_{na}|^2 = \sum_b |u_{nb}|^2 = 1$. The average phase derivative weighted by intensity over the outgoing channels $b$ is thus $<\frac{d\varphi_{nb}^{(v)}}{d\omega}>_b = \sum_b |u_{nb}|^2 \frac{d\varphi_{nb}}{d\omega} = \frac{1}{i} \mathbf{u}_n^* \cdot \frac{d\mathbf{u}_n}{d\omega}$. Using Eq. (2), $<\frac{d\varphi_{nb}^{(v)}}{d\omega}>_b$ is equal to the time shift of the barycenter of $v_{nb}(t)$ averaged over channels $b$,

$$\frac{1}{i}\mathbf{u}_n^* \cdot \frac{d\mathbf{u}_n}{d\omega} = \lim_{\Delta\omega \to 0} <\frac{\int |u_{nb}(t;\omega_0,\Delta\omega)|^2 t\, dt}{\int |u_{nb}(t;\omega_0,\Delta\omega)|^2 dt}>_b \quad (13)$$

The relation between the transmission delay time and the $d\theta_n/d\omega$ is then,
$$\lim_{\Delta\omega \to 0} \Delta t_n(\omega_0, \Delta\omega) = d\theta_n/d\omega \quad (14)$$

This relation is verified by the agreement of the spectra of $\Delta t_n$ and $d\theta_n/d\omega$ in Fig. S3. Note that the demonstration presented here regarding the equality of $d\theta_n/d\omega$ and the transmission eigenchannel delay time does not involve any assumption regarding the degree of completeness of the measured TM. Equation (14) holds as well in the case of a lossy medium.

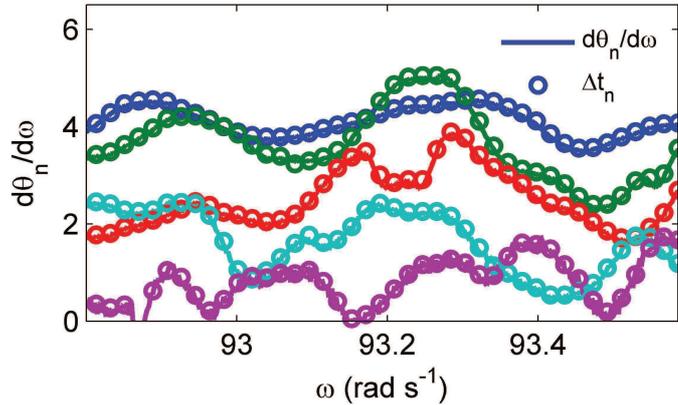

**Figure S3 | Measured spectra of eigenchannel phase derivative and delay time in a single realization of the random sample.** The spectra for eigenchannels $n$=1,5,9,13,17,21,25 of $d\theta_n/d\omega$ for a sample of diffusive waves with g=6.9 (lines) are compared to the delay times of the corresponding eigenchannels found with Eq. (12) (circles).

4. **Tail of the distribution of the DOS for localized waves**

In the main text, the distribution of the measured DOS for localized waves and $L_{eff}/\xi$=2.08 is seen to exhibit an algebraic tail as $1/\rho^{4.8}$. This is reminiscent of one-dimensional theoretical studies showing that the distribution of the DOS has an algebraic tail as $1/\rho$ for localized waves and $L \to \infty$ [17, 18]. In Fig. S4, we show that this scaling of the tail is in agreement with computer simulations. $P(\rho)$ is also seen to have an algebraic tail, with $P(\rho) \propto 1/\rho^4$ for $\rho \gg <\rho>$.

In our multichannel system the contribution of the second eigenchannel to the EDOS cannot be neglected for $L/\xi$ not much larger than one. We attribute the difference of the exponent giving the best fit to this contribution as well as the finite value of $L$.

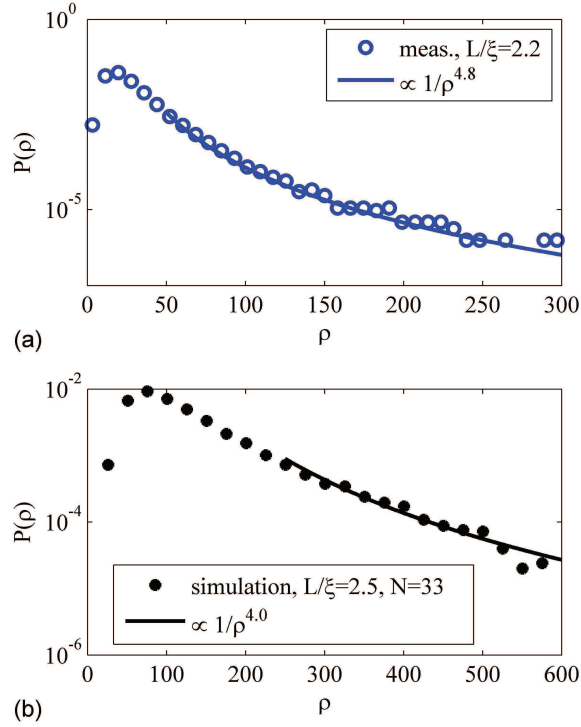

**Figure S4 | Probability distribution of the DOS for localized waves.** (a) Measurement of the probability distribution of the DOS for $L_{eff}/\xi$=2.08 (blue circles). The blue line is an algebraic tail as $1/\rho^{4.8}$. (b) Simulations of the probability distribution of the DOS for $L/\xi$=2.5 and N=33. The black line is an algebraic tail as $1/\rho^4$.

5. **Relation between the EDOS and the transmission eigenvalues for localized waves**

The variation of the transmission delay time normalized by the average single channel delay time $\langle d\theta_n/d\omega \rangle/\langle d\varphi/d\omega \rangle$ with the transmission eigenvalue in diffusive samples is seen to collapse to a single curve for different values of $L/\xi$ in Fig. 5c of the main text. For localized waves, only a single eigenchannel contributes significantly to transmission. To explore the variation of $d\theta_n/d\omega$ with transmission, we consider sub-ensembles of $d\theta_n/d\omega$ for fixed values of $\tau_n$. The average of those sub-ensembles is denoted by $<d\theta_n/d\omega>_\tau$ for transmission $\tau$. $<d\theta_n/d\omega>_\tau/<d\varphi/d\omega>$ is seen in Fig. S5 to increase with $<\tau>$ for localized waves $L/\xi>1$ as was the case for diffusive waves for $L/\xi<1$. This reflects that high transmission is associated with deeper penetration into the sample. The integral of the intensity profile is therefore higher for more strongly transmitting eigenchannels and along with this the DOS is higher.

The curves for normalized delay time $<d\theta_n/d\omega>_\tau/<d\varphi/d\omega>$ vs. $<\tau>$ do not overlap for localized waves, as was the case for diffusive waves; rather the normalized delay time increases with $L/\xi$ for the a given value of transmission. Because the wave is transmitted through sharp resonances for larger values of $L/\xi$, fluctuations of the EDOS are expected to be larger than for diffusive waves and the delay time in transmission normalized by its average can reach much higher values when the wave is on resonance with a mode and transmission is high.

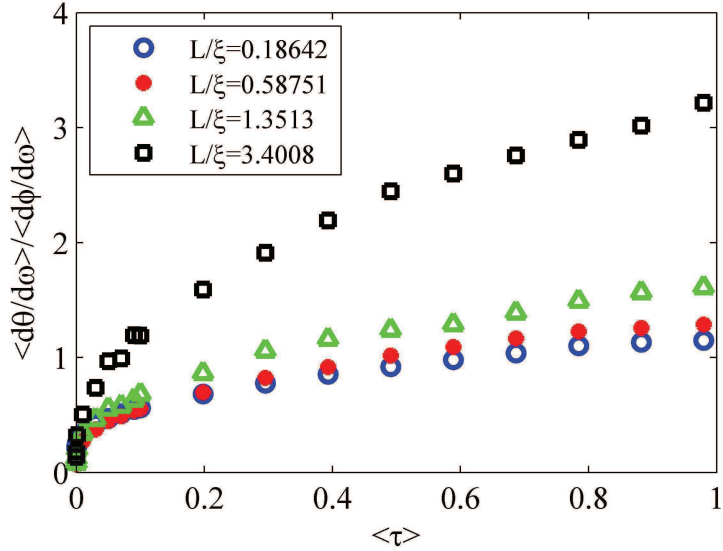

**Figure S5 | Simulations of the variation of the EDOS with transmission for localized waves.** Simulations for the complete TM give $<d\theta_n/d\omega>_\tau/<d\varphi/d\omega>$ with for $L/\xi$=0.18 (blue circles), $L/\xi$=0.58 (red dots), $L/\xi$=1.35 (green triangles) and $L/\xi$=3.4 (black squares).

6. **Incomplete measurement of the transmission matrix**

It is not possible to capture all the transmitted energy in all the outgoing channels and excite the sample with all possible incident channels. In optics, the incompleteness of the measurement of TM of a disordered slab is due to the finite size of the illumination beam, the finite numerical aperture of the collecting lens and loss of energy because of the scattered photons escaping from the sides of the sample. In our microwave measurements, only a fraction of the transmitted energy is measured since the TM is determined by measuring the transmission coefficients on a grid with a finite number of points [8]. The impact of the incomplete measurement of the TM upon the probability density of eigenvalues of the TM has been recently investigated by Goetschy and Stone [19]. They explored the change in the density of transmission eigenvalues from the bimodal distribution to a distribution characteristics of Gaussian random matrices as the ratio of measured channels $N'$ and total number of channels $N$ on the input side $m_1=N_1'/N$ and output side $m_2=N_2'/N$ decreases. Since the spacing of the grids on which the source and detector are moved is the same, we assume then $m_1=m_2$. The density of the transmission eigenvalues for the incomplete TM depends only upon on the value of $m$ and the ratio of the transport mean free path $l$ and the sample length $L$, which is close to the value of average total transmission $<T_a>$. Here $a$ indicates the particular incident channel or position of the detector. In our sample, the value of $<T_a>$ is $g/N$= 0.115. The best agreement between measured probability distribution of the eigenvalues in our diffusive samples of length $L$ = 23 cm and the theoretical calculation is obtained for a value of $m = 0.7$. This is consistent with the degree of control reported in the main text by comparing the measured $\langle d\theta_n/d\omega\rangle/\langle d\phi_{ba}/d\omega\rangle$ and simulations in which only 70% of the channels are used.

**References**

[1] M. G. Krein, Sov. Math. Dokl **3**, 707 (1962).